\documentclass[floatfix,aps,prl,twocolumn,showpacs,superscriptaddress,preprintnumbers]{revtex4-1}

\usepackage[english]{babel}
\usepackage{bm}    
\usepackage{epsfig}
\usepackage{amsbsy,amssymb} 
\usepackage{graphicx}
\usepackage{array}
\usepackage{amsmath}
\usepackage{multirow}
\usepackage{natbib}

\newcommand\new{\newcommand}         

\def\beq{\begin{equation}}   
\def\eeq{\end{equation}}
\def\bea{\begin{eqnarray}}  
\def\eea{\end{eqnarray}} 
\newcommand{\bite}{\begin{itemize}}
\newcommand{\eite}{\end{itemize}}

\new{\eV}         {{\ifmmode {\mathrm{ eV}}\else ${\mathrm{ eV}}$\fi}}
\new{\MeV}        {{\ifmmode {\mathrm{ MeV}}\else ${\mathrm{ MeV}}$\fi}}
\new{\MeVc}       {{\ifmmode {\mathrm{ MeV}}/c\else ${\mathrm{ MeV}}/c$\fi}}
\new{\MeVcc}      {{\ifmmode {\mathrm{ MeV}}/c^2\else ${\mathrm{ MeV}}/c^2$\fi}}
\new{\GeV}        {{\ifmmode {\mathrm{ GeV}}\else ${\mathrm{ GeV}}$\fi}}
\new{\GeVc}       {{\ifmmode {\mathrm{ GeV}}/c\else ${\mathrm{GeV}}/c$\fi}}
\new{\GeVcc}      {{\ifmmode {\mathrm{ GeV}}/c^2\else ${\mathrm{GeV}}/c^2$\fi}}
\new{\TeV}        {{\ifmmode {\mathrm{ TeV}}\else ${\mathrm{ TeV}}$\fi}}

\new{\Mh}         {{\ifmmode M_{\mathrm{ H}}
                    \else $M_{\mathrm{H}}$\fi}}
\new{\Mz}         {{\ifmmode M_{\mathrm{Z}}
                    \else $M_{\mathrm{Z}}$\fi}}
\new{\Mzsq}       {{\ifmmode M^2_{\mathrm{ Z}}
                    \else $M^2_{\mathrm{Z}}$\fi}}
\new{\as}[1]      {{\ifmmode\alpha^{#1}_s
                    \else$\alpha^{#1}_s$\fi}}
\new{\asx}[1]      {{\ifmmode a^{#1}_s
                    \else $a^{#1}_s$\fi}}
\new{\asb}[1]     {{\ifmmode\overline{\alpha}^{#1}_s
                    \else $\overline{\alpha}^{#1}_s$\fi}}
\new{\asmz}       {{\ifmmode\alpha_s(\Mzsq)
                    \else $\alpha_s(\Mzsq)$\fi}}
\new{\lqcd}       {{\ifmmode\Lambda_{\mathrm{ QCD}}
                    \else $\Lambda_{\mathrm{ QCD}}$\fi}}

\def\Gosam{{{\sc GoSam}}}

\def\samurai{{{\sc samurai}}}
\def\Sherpa{{{\sc Sherpa}}}
\def\Amegic{{{\sc Amegic}}}

\def\C++{{{\sc c++}}}

\def\QCDLoop{{{\sc QCDLoop}}}
\def\OneLoop{{{\sc OneLoop}}}
\def\Golem{{{\sc Golem95C}}}

\newcommand{\ci}{g_{\mbox{\tiny eff}}}

\begin{document}

\title{NLO QCD corrections to Higgs boson production plus three jets in gluon fusion}

\author{G.~Cullen}
\affiliation{Deutsches Elektronen-Synchrotron DESY, Platanenallee 6, 15738 Zeuthen, Germany}
\author{H.~van Deurzen}
\affiliation{Max-Planck-Institut f\"ur Physik, F\"ohringer Ring 6,
80805 M\"unchen, Germany}
\author{N.~Greiner}
\affiliation{Max-Planck-Institut f\"ur Physik, F\"ohringer Ring 6,
80805 M\"unchen, Germany}
\author{G.~Luisoni}
\affiliation{Max-Planck-Institut f\"ur Physik, F\"ohringer Ring 6,
80805 M\"unchen, Germany}
\author{P.~Mastrolia}
\affiliation{Max-Planck-Institut f\"ur Physik, F\"ohringer Ring 6,
80805 M\"unchen, Germany}
\affiliation{Dipartimento di Fisica e Astronomia, Universit\`a di
Padova, and INFN
Sezione di Padova, via Marzolo 8, 35131 Padova, Italy}
\author{E.~Mirabella}
\affiliation{Max-Planck-Institut f\"ur Physik, F\"ohringer Ring 6,
80805 M\"unchen, Germany}
\author{G.~Ossola}
\affiliation{New York City College of Technology, City University of
New York, 300 Jay Street, Brooklyn NY 11201, USA}
\affiliation{The Graduate School and University Center, City
University of New York, 365 Fifth Avenue, New York, NY 10016, USA}
\author{T.~Peraro}
\affiliation{Max-Planck-Institut f\"ur Physik, F\"ohringer Ring 6,
80805 M\"unchen, Germany}
\author{F.~Tramontano}
\affiliation{Dipartimento di Fisica, Universit\`a degli
studi di Napoli ``Federico II'', I-80125 Napoli, Italy}
\affiliation{INFN, Sezione di Napoli, I-80125 Napoli, Italy}


\preprint{DESY-13-119, DF-06-2013, LPN13-042, MPP-2013-193, SFB/CPP-13-46}
\begin{abstract}
We report on the calculation of the cross section for Higgs boson
production in association with three jets via gluon fusion, at
next-to-leading-order (NLO) accuracy in QCD, in the infinite top-mass
approximation. After including the complete NLO QCD corrections, we
observe a strong reduction in the scale dependence of the result, and
an increased steepness in the transverse momentum distributions of
both the Higgs and the leading jets. The results are obtained with the
combined use of {\Gosam}, {\Sherpa}, and the {\sc
  MadDipole/MadEvent} framework.
\end{abstract}

 

\maketitle

\section{Introduction}
\label{Sec:intro}

The latest results reported by the ATLAS and CMS collaborations have
confirmed with a higher confidence-level the existence of a new
neutral boson with mass of about $125-126$ {\GeV} and spin different
from one~\cite{Aad:2012tfa,Chatrchyan:2012ufa}, and suggest that the
new particle has indeed the features of a Higgs boson, thus confirming
the validity of the electroweak symmetry breaking mechanism.
Although the evidence accumulated so far is compatible with the
hypothesis that the new resonance is the Higgs particle predicted by the
Standard Model (SM) with the $J^P=0^+$~\cite{ATLAS:2013mla, CMS:yva},
in order to confirm its nature, further high-precision studies on
spin, parity, coupling strengths and branching ratios are mandatory.

In $pp$-collisions, the dominant Higgs production mechanism proceeds
via gluon fusion (GF), $gg \to H$, where the coupling of the Higgs to
the gluons is mediated by a heavy quark loop.

Another important production channel for the Higgs boson is Vector
Boson Fusion (VBF), since it allows a direct measurement of the
coupling of the Higgs to the massive electroweak
bosons~\cite{Zeppenfeld:2000td}. The cross section in the VBF channel
is about an order of magnitude smaller than in GF, and even after
applying specific cuts, the latter remains the main source of
background for Higgs production in VBF.

For these reasons, the calculation of higher order corrections for the
GF production of a Higgs boson in association with jets has received a
lot of attention in the theory community over the past
decades~\cite{Dittmaier:2011ti, Dittmaier:2012vm, Heinemeyer:2013tqa}.

The leading order (LO) contribution to the production of a Higgs boson
in association with two jets ($Hjj$), and three jets ($Hjjj$) have
been computed respectively in
Refs.~\cite{DelDuca:2001eu,DelDuca:2001fn}, and in the recent
Ref.~\cite{Campanario:2013mga}.  These calculations have been
performed retaining the full top-mass ($m_t$) dependence, and showed
the validity of the large top-mass approximation ($m_{t} \to \infty$)
whenever the mass of the Higgs particle and the $p_T$ of the jets are
not sensibly larger than the mass of the top quark.
In this approximation, the Higgs coupling to two gluons, which at LO
is mediated by a top-quark loop, becomes independent of $m_t$, and it
can be described by an effective operator~\cite{Wilczek:1977zn}, as
\bea
\mathcal{L}_{\rm eff} =- \frac{\ci}{4} \, H \, \mbox{tr} \left(G_{\mu \nu} G^{\mu \nu} \right ) \, .
\label{Eq:EffL}
\eea 
In the $\overline{\mbox{MS}}$ scheme, the coefficient $\ci$
reads~\cite{Djouadi:1991tka,Dawson:1990zj} 
\bea
\ci = -\frac{\alpha_s}{3 \pi v} \left ( 1 + \frac{11}{4 \pi} \alpha_s\right )+ \mathcal{O}(\alpha_s^3)\,,
\eea
in terms of the Higgs vacuum expectation value $v$, set to $v=246$ GeV.  The
operator~(\ref{Eq:EffL}) leads to new Feynman rules, with vertices
involving the Higgs field and up to four gluons.

\begin{figure}[t]
\begin{center}
\includegraphics[width=2.5cm]{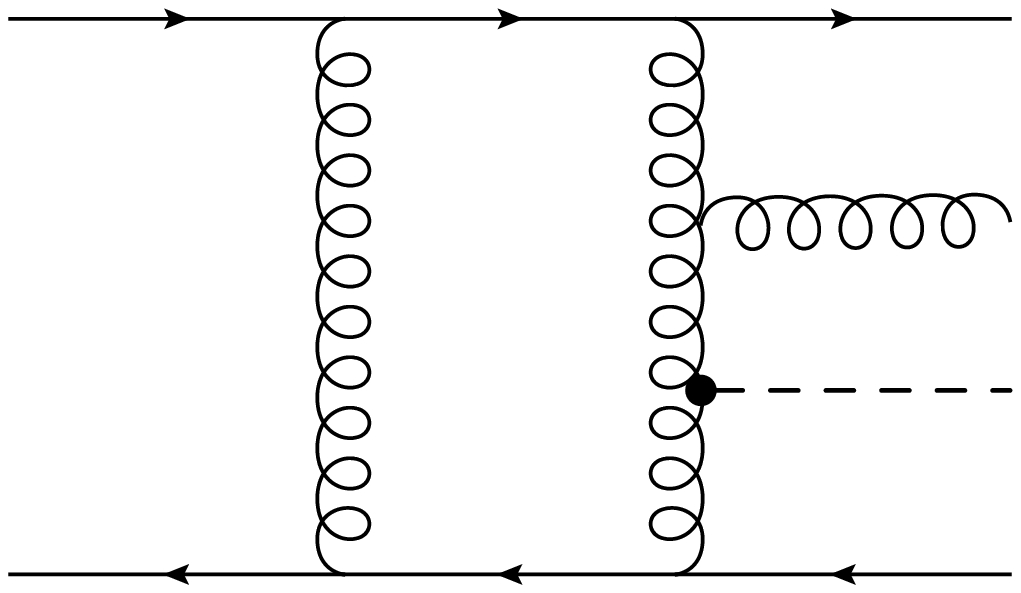} \qquad \quad
\includegraphics[width=2.5cm]{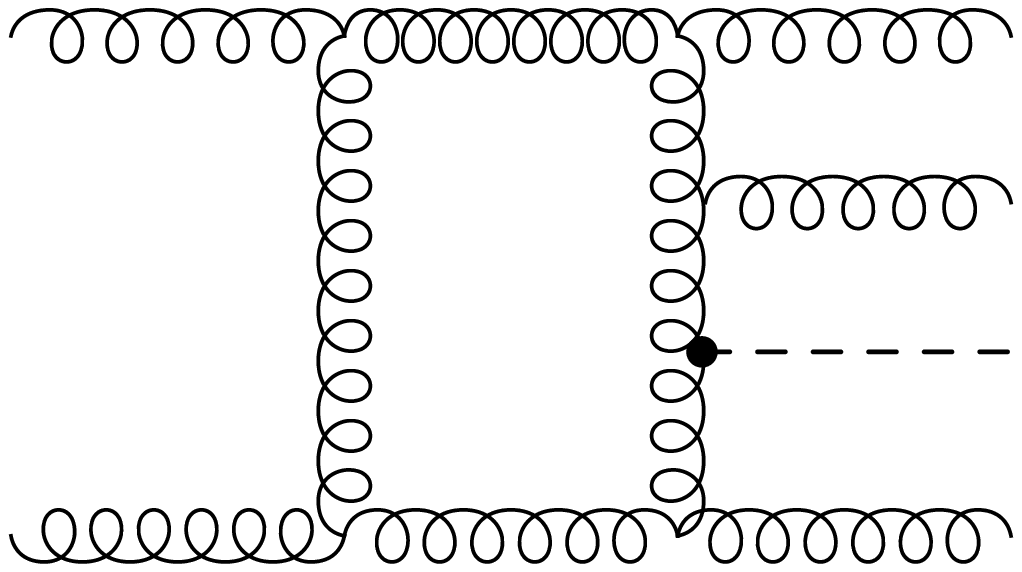} 
\caption{Sample hexagon diagrams which enter in the six-parton
  one-loop amplitudes for $q {\bar q} \to H q {\bar q} g $ and $gg \to
  H g g g$. The dot represents the effective $g g H$ vertex.}
\label{Fig:exagons}
\end{center}
\end{figure}

The leading order contributions to $Hjjj$, both for VBF and GF (in the
$m_t \to \infty$ limit), have been calculated
in~\cite{DelDuca:2004wt}.  However, while the VBF calculation is
available also at NLO~\cite{Figy:2007kv}, the computation of the Higgs
plus three jets in GF is still missing.

Elaborating on the techniques employed in the recent calculation of
the NLO contributions to $Hjj$ production at the
LHC~\cite{vanDeurzen:2013rv}, in this letter we report on the
calculation of the cross section for  $p p \to H j j j $ in GF at NLO
accuracy in QCD, within the infinite $m_t$ approximation.

This calculation is challenging due to the complexity of both the
real-emission contributions and of the virtual corrections, which
involve more than $10,000$ one-loop Feynman diagrams with up to
rank-seven hexagons.

\section{Computational setup}
\label{sec:calc}

A complete next-to-leading order calculation 
requires the evaluation of virtual and real emission
contributions.

For the computation of the virtual corrections we use a code generated
by the program package {\Gosam}~\cite{Cullen:2011ac}, which combines
automated diagram generation and algebraic manipulation
\cite{Nogueira:1991ex, Vermaseren:2000nd, Reiter:2009ts,
  Cullen:2010jv} with integrand-level reduction techniques
\cite{Ossola:2006us, Ossola:2007bb, Ellis:2007br,
  Ossola:2008xq,Mastrolia:2008jb, Mastrolia:2012bu,Mastrolia:2012an}.

In order to deal with the complexity level of the considered
calculation, the {\sc GoSam} code has been enhanced. On the one side,
the generation algorithm has been improved by a more efficient
diagrammatic layout: Feynman diagrams are grouped according to their
topologies, namely global numerators are constructed by combining
diagrams that have a common set, or subset, of denominators,
irrespectively of the specific particle content.  On the other side,
additional improvements in the performances of {\sc GoSam} have been
achieved by exploiting the optimized manipulation of polynomial
expressions available in {\sc Form 4.0}~\cite{Kuipers:2012rf}.  The
new developments of {\sc GoSam}, regarding the improved generation and
reduction algorithms, will be properly discussed in a dedicated
communication.

Within the {\sc GoSam} framework the virtual corrections are evaluated
using the $d$-dimensional integrand-level decomposition implemented in
the \samurai\ library~\cite{Mastrolia:2010nb,Mastrolia:2012du}, which
allows for the combined determination of both cut-constructible and
rational terms at once.
Alternatively, a tensorial decomposition
\cite{Binoth:2008uq,Heinrich:2010ax} via {\sc Golem95} is used as a
rescue system.  After the reduction, all relevant master integrals are
computed by means of {\QCDLoop}~\cite{vanOldenborgh:1990yc,
  Ellis:2007qk}, {\OneLoop}~\cite{vanHameren:2010cp}, or
{\Golem}~\cite{Cullen:2011kv}. 

The basic partonic processes contributing to $Hjjj$ production are
listed in Tab.~\ref{Tab:sub}, together with the corresponding number
of Feynman diagrams and the approximate computing time per
phase-space point after summing over color and helicities.
Representative one-loop diagrams are depicted in
Figure~\ref{Fig:exagons}.

\begin{table}[h]
\begin{tabular}{ l  c  c } 
\hline  
\hline
{\sc Subprocess} & \quad \qquad {\sc Diagrams}  \qquad \quad & {\sc Time/PS-point} [sec] \\
\hline
$q \bar q \to H q' \bar q' g$ &  467 &  0.29 \\
$q \bar q \to H q  \bar q  g$ &  868 &  0.60 \\
$g      g \to H q  \bar q  g$ & 2519 &  3.9  \\
$g      g \to H g       g  g$ & 9325 &  20   \\
\hline
\hline
\end{tabular}
\caption{Number of Feynman diagrams and computing time per phase-space
  point for each subprocess, on a Intel i7 960 (3.20GHz) CPU. The code
  is compiled with the Intel fortran compiler {\tt ifort} (with
  optimization {\tt O2}).}
\label{Tab:sub}
\end{table}

The ultraviolet (UV), the infrared (IR), and the collinear
singularities are regularized using dimensional reduction (DRED).  UV
divergences have been renormalized in the $\overline{\mbox{MS}}$
scheme.  In the case of LO [NLO] contributions we describe the running
of the strong coupling constant with one-loop [two-loop] accuracy.

The effective $Hgg$ coupling leads to integrands that may exhibit
numerators with rank larger than the number of the denominators.  In
general, for these cases, the parametrization of the residues at the
multiple-cut has to be extended and, as a consequence, the
decomposition of any one-loop amplitude acquires new master integrals
(MIs)~\cite{Mastrolia:2012bu}.  The extended integrand decomposition
has been implemented in the \samurai\ library.

Remarkably, for the processes at hand, it has been proven that the
higher-rank terms are proportional to the loop momentum squared, which
simplifies against a denominator, hence generating lower-point
integrands where the rank is again equal to the number of
denominators~\cite{vanDeurzen:2013rv}. Consequently, the coefficients
of the new MIs have to vanish identically, as explicitly verified.
The available options in {\Gosam} for the algebraic manipulation of
the integrands allow for the automatic computation of the virtual
corrections in two different ways. In the first approach, {\Gosam}
decomposes the four-dimensional part of the numerators using the
extended-rank decomposition, and adds the analytic results of the
rational terms (generated from the extra-dimensional part). In the
second approach, the regular decomposition of {\samurai}, without the
higher rank extension, is employed on the whole $d$-dimensional
integrands. We checked that both approaches provide identical
answers. In the following, we adopt the second strategy, which proved
to be numerically more efficient.

The double and the single poles conform to the universal singular
behavior of dimensionally regulated one-loop
amplitudes~\cite{Catani:2000ef}.  We also checked that our results
fulfill gauge invariance: when substituting the polarization vectors of one or more
gluons with the corresponding momenta, the
result for the amplitudes, after summing over all diagrams, are indeed
vanishing.
Additional information about the virtual contributions can be found in the
Appendix.

Results for the cross section are obtained with a hybrid setup which
combines the features of two different Monte Carlo (MC) tools. For the
generation and integration of the Born and of the virtual
contributions, we used an automated framework for fixed order NLO QCD
calculations, based on the interplay of {\Gosam} and
{\Sherpa}~\cite{Gleisberg:2008ta}, where the tree-level matrix
elements are obtained with the \Amegic~\cite{Krauss:2001iv}
library. The integration is carried out by generating
$\mathcal{O}(10^6)$ events, sampled on a MC grid trained on the Born
matrix element, and weighted with the sum of the Born and the virtual
amplitudes.

For the integration of the real-radiation terms, the
dipole-subtraction terms, and the integrated dipoles, we employ a
combination of {\sc MadGraph}~\cite{Stelzer:1994ta,Alwall:2007st}
(matrix elements), {\sc
  MadDipole}~\cite{Frederix:2008hu,Frederix:2010cj} (subtraction
terms), and {\sc MadEvent}~\cite{Maltoni:2002qb} (numerical
integration). We verified the independence of our result under the
variation of the so called $\alpha$-parameter that fixes the amount of
subtractions around the divergences of the real corrections.

We first proved the consistency of our hybrid MC integration on $pp
\to Hjj$, verifying that the full cross section at NLO agrees with the
corresponding result for the integration of both the virtual and the real
corrections obtained by the interplay of {\Sherpa} and {\Gosam} alone.  Moreover, for the process
under consideration, namely $pp \to Hjjj$, we found excellent
agreement between {\sc MadGraph} and {\Sherpa} for the LO 
cross section.

\section{Integrated Cross section}

In the following, we present results for the integrated
cross section of Higgs boson plus three jets production
at the LHC, for a center-of-mass energy of $8$ TeV.
The mass of the Higgs boson is set to $m_H=125$ GeV.

Jets are clustered using the {\tt antikt}-algorithm implemented in
{\sc FastJet}~\cite{Cacciari:2005hq,Cacciari:2008gp,Cacciari:2011ma}
with radius $R=0.5$ and a minimum transverse momentum of
$p_{T,jet}>20$ GeV and pseudorapidity $|\eta|<4.0$. The LO cross
section is computed with the LO parton-distribution functions {\tt cteq6L1},
whereas at NLO we use {\tt cteq6mE}~\cite{Pumplin:2002vw}.

\begin{figure}[h]
\begin{center}
\includegraphics[width=8.0cm]{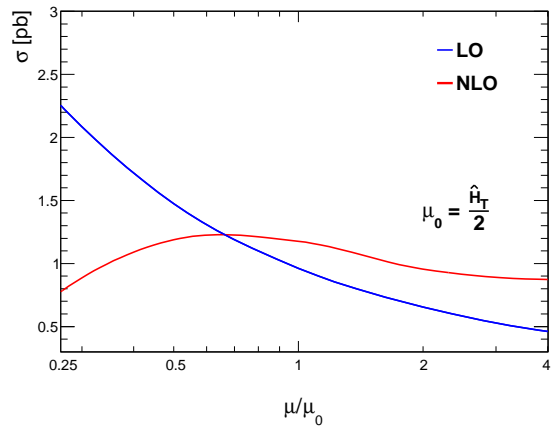} 
\caption{Scale dependence of the total cross section at LO and NLO.}
\label{Fig:xs_scale_dep}
\end{center}
\end{figure}

Everywhere, but in the effective coupling of the Higgs to the gluons,
the renormalization and factorization scales are set to 
\beq
\mu_{F}=\mu_{R}=\frac{\hat{H}_T}{2}=\frac{1}{2}\left(\sqrt{m_{H}^{2}+p_{T,H}^{2}}+\sum_{i}|p_{T,i}|\right)\,,
\eeq 
where the sum runs over the final state jets. The strong coupling
is therefore evaluated at different scales according to
$\alpha_s^5\rightarrow\alpha_s^2(m_H)\alpha_s^3(\hat{H}_T/2)$.  The
theoretical uncertainties are estimated by varying the scales by
factors of $0.5$ and $2.0$ respectively.
In the effective coupling the scale is kept at $m_H$. 
Within this setup we obtain the following total cross section at
LO and NLO:
$$\sigma_{\rm LO}[{\rm pb}] = 0.962^{+0.51}_{-0.31} \ ,\quad \sigma_{\rm NLO}[{\rm pb}]= 1.18^{+0.01}_{-0.22} \ .$$ 
The scale dependence of the total cross section, depicted in
Fig.~\ref{Fig:xs_scale_dep}, is strongly reduced by the inclusion of
the NLO contributions.

\begin{figure}[h]
\begin{center}
\includegraphics[width=8.0cm]{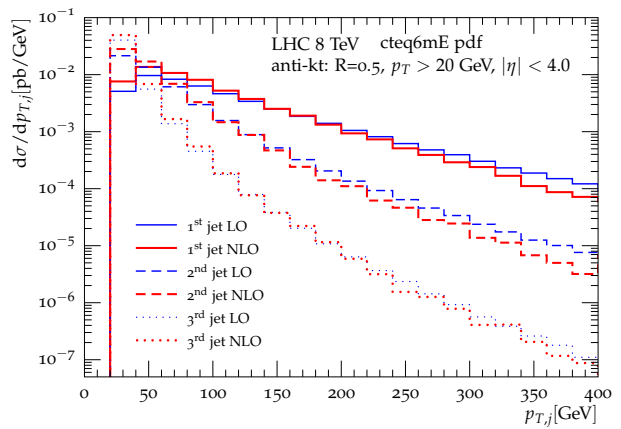} 
\caption{Transverse momentum ($p_T$) distributions for the first,
  second, and third leading jet.}
\label{Fig:jets}
\end{center}
\end{figure}

\begin{figure}[h]
\begin{center}
\includegraphics[width=8.0cm]{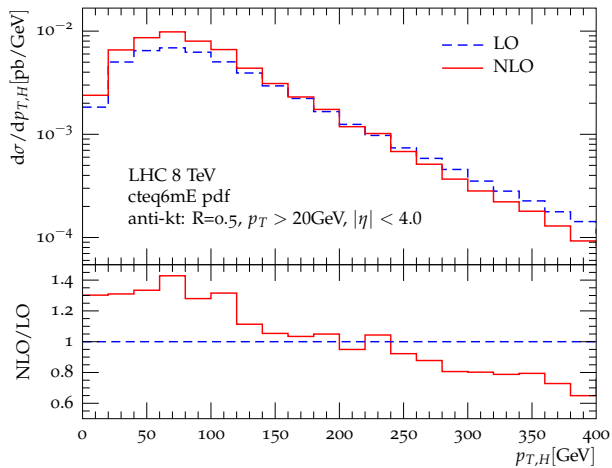} 
\caption{Transverse momentum ($p_T$) distributions for the Higgs boson.}
\label{Fig:higgs}
\end{center}
\end{figure}

In Figs.~\ref{Fig:jets} and~\ref{Fig:higgs}, we show the $p_T$
distributions of the three jets and of the Higgs boson,
respectively. The NLO corrections enhance all distributions for $p_T$
values lower than $150-200$ GeV, whereas their contribution is
negative at higher $p_T$. This behavior is explicitly shown in the
lower part of Fig.~\ref{Fig:higgs} for the case of the Higgs boson.

This study also shows that the virtual contributions for $pp \to Hjjj$
generated by {\Gosam} can be successfully paired with available Monte
Carlo programs to aim at further phenomenological analyses.


\bigskip

\begin{acknowledgments}
We thank Thomas Hahn and Gudrun Heinrich for discussions and comments
on the manuscript, and Marek Sch\"onherr for assistance with the usage
of \Sherpa. The work of G.C. was supported by DFG SFB-TR-9 and the EU
TMR Network LHCPHENOnet.  The work of H.v.D., G.L., P.M., and
T.P. was supported by the Alexander von Humboldt Foundation, in the
framework of the Sofja Kovaleskaja Award 2010, endowed by the German
Federal Ministry of Education and Research.  G.O. was supported in
part by the National Science Foundation under Grant
PHY-1068550. F.T. acknowledges partial support by MIUR under project
2010YJ2NYW.  G.C. and G.O. wish to acknowledge the kind hospitality of
the Max-Planck-Institut f\"ur Physik in Munich at several stages
during the completion of this project. This research used computing
resources from the Rechenzentrum Garching and the New York City
College of Technology.
\end{acknowledgments}

\section*{Appendix: Selected Results for the Virtual Contributions}

The numerical values of the one-loop sub-amplitudes, defined as
 \bea \frac{ 2\, \mathfrak{Re} \left \{
  \mathcal{M}^{\mbox{\tiny tree-level} \ast } \mathcal{M}^{\mbox{\tiny
      one-loop} } \right \} }{( \alpha_s / 2 \pi) \left |
  \mathcal{M}^{\mbox{\tiny tree-level}} \right |^2 } \equiv
\frac{a_{-2}}{\epsilon^2} + \frac{a_{-1}}{\epsilon} + a_0 \, ,
\label{Eq:AI}
\eea and evaluated at the non-exceptional phase space point given in
Tab.~\ref{Tab:ppsJ3}, are collected in Tab.~\ref{Tab:resJ3}.  The
values of the double and the single poles conform to the universal
singular behavior of dimensionally regulated one-loop
amplitudes~\cite{Catani:2000ef}. The precision of the finite parts is
estimated by re-evaluating the amplitudes for a set of momenta rotated
by an arbitrary angle about the axis of collision.

\begin{table*}[ht]
\centering
\begin{tabular}{c c c c c}
\hline 
\hline 
{\sc particle} & $E$& $p_x$ & $p_y$ & $p_z$ \\ 
\hline
$p_1$ & 250.00000000000000 & 0.0000000000000000 & 0.0000000000000000 & 250.00000000000000 \\
$p_2$ & 250.00000000000000 & 0.0000000000000000 & 0.0000000000000000 &-250.00000000000000 \\
$p_3$ &	131.06896655823209 & 27.707264814722667 &-13.235482900394146 & 24.722529472591685 \\
$p_4$ & 164.74420140597425 &-129.37584098675183 &-79.219260486951597 &-64.240582451932028 \\
$p_5$ & 117.02953632773803 & 54.480516624273569 & 97.990504664150677 &-33.550658370629378 \\
$p_6$ & 87.157295708055642 & 47.188059547755266 &-5.5357612768047906 & 73.068711349969661 \\
\hline
\hline  
\end{tabular}
\caption{Benchmark phase space point for Higgs plus three jets
  production. Particles are ordered as in Tab.~\ref{Tab:sub}.}
\label{Tab:ppsJ3}
\end{table*}

\begin{table*}[ht]
\begin{tabular}{ l  r  r  r  r } \hline \hline 
         & $g g \to H g g g$  & $g g \to H q \bar q g$   &  $q \bar q \to H   q \bar q g$    &  $q \bar q \to   H q' \bar q' g$    \\  
\hline  
$a_0$     \qquad &   \underline{41.2287}8766741685                &   \underline{48.6842413}4989478 
  &    \underline{ 69.3235114047}4695   &  \underline{15.79262767}177915    \\
$a_{-1}$  \qquad &   \underline{ -47.16715}419132659              & \underline{ -36.08277728}077228 
  &  \underline{  -29.988629329636}59  &  \underline{ -32.35320587073}968  \\
$a_{-2}$  \qquad &   \underline{ -14.9999999999999}1             & \underline{ -11.666666666666}83   
  &  \underline{ -8.33333333333333}9  &  \underline{  -8.3333333333333}98  \\
\hline
\hline 
\end{tabular}
\caption{Numerical results for the four subprocesses listed in
  Tab.~\ref{Tab:sub} evaluated at the phase space point of
  Tab.~\ref{Tab:ppsJ3}.  The accuracy of the result is indicated by
  the underlined digits.
} 
\label{Tab:resJ3}
\end{table*}

\begin{figure}[h]
\begin{center}
\includegraphics[width=8.0cm]{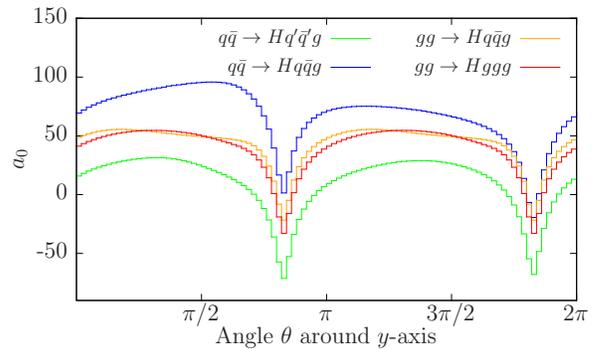} 
\caption{Finite-term $a_0$ of the virtual matrix-elements for $q {\bar q}
  \to H q' {\bar q'} g$ (green), $q {\bar q} \to H q {\bar q} g $
  (blue), $gg \to H q {\bar q} g $ (orange), $gg \to H g g g$ (red).}
\label{Fig:Hjjj}
\end{center}
\end{figure}

In Fig.~\ref{Fig:Hjjj}, we present the results for the finite part
$a_0$ of the virtual matrix elements for the various subprocesses
calculated along a certain one-dimensional curve in the space of final
state momenta.
Starting from the phase space point in Tab.~\ref{Tab:ppsJ3}, in which
the initial partons lie along the $z$-axis, we generate new
configurations by rotating the final state momenta by an angle $\theta
\in [0, 2\pi ] $ about the $y$-axis.

\bibliographystyle{apsrev}
\bibliography{references.bib}

\end{document}